\documentclass[preprint,12pt]{elsarticle}

\usepackage{amssymb}

\journal{PSS Special Issue: Outer Planets VIII}

\begin{document}

\begin{frontmatter}



\title{Dust cloud lightning in extraterrestrial atmospheres}


\author{Christiane Helling$^1$, Moira Jardine$^1$, Declan Diver$^2$, S\"oren Witte$^3$}

\address{
$^1$ SUPA, School of Physics and Astronomy, University of St Andrews, St Andrews, KY16 9SS, UK,\\
 $^2$ School of Physics and Astronomy, University of Glasgow, Glasgow G12 8QQ, UK\\
$^3$ Hamburger Sternwarte, Gojenbergsweg 112, 21029 Hamburg, Germany}

\begin{abstract}
Lightning is present in all solar system planets which form clouds in
their atmospheres.  Cloud formation outside our solar system is
possible in objects with much higher temperatures than on Earth or on
Jupiter: Brown dwarfs and giant extrasolar gas planets form clouds
made of mixed materials and a large spectrum of grain sizes.  These
clouds are globally neutral obeying dust-gas charge equilibrium which
is, on short timescales, inconsistent with the observation of
stochastic ionization events of the solar system planets. We argue
that a significant volume of the clouds in brown dwarfs and extrasolar
planets is susceptible to local discharge events and that the upper
cloud layers are most suitable for powerful lightning-like discharge
events. We discuss various sources of atmospheric ionisation, including thermal ionisation and a first estimate of  ionisation by cosmic rays, and argue
that we should expect thunderstorms also in the atmospheres of
brown dwarfs and giant gas planets which contain mineral clouds.
\end{abstract}


\end{frontmatter}

\section{Introduction}
Until recently, clouds were believed to be unique to Earth-like planets and to the gas giant
planets that are rather far away from their host star (like Jupiter,
Saturn, Uranus in our solar system). Extrasolar planets are now a
matter of fact and their diversity has increased over the last couple
of years due to various ground based observational efforts like
SuperWASP, HAT, TrES and regarding giant gas planets, and by the CoRot
and the Kepler space mission with respect to Earth-like, low-mass
planets. Compared to the solar system, however, many giant gas planets
are orbiting their host star at very short distance.  Observations
have revealed that hazes appear in the upper atmospheres of such
close-in planets, because the haze absorbing the stellar radiation
during transit makes the planet appear larger than expected. The
transit spectroscopy of HD 189733b presented in Pont et al. (2008) and
in Sing et al. (2011) provides the first proof that small mineral
particles do not only populate the highest layers of the terrestrial
atmospheres but are also present in extrasolar Jupiters.  Such direct
observations of atmospheric dust have not yet been possible for brown
dwarfs.  Brown dwarfs have the same size and effective temperature as
the gas-giants, and they have been subject to extensive direct
spectroscopic observations as they are much more close by and, hence,
it is easier to take direct spectroscopic measurements for brown
dwarfs than for the majority of the exoplanets. High- and
low-resolution spectra, reaching from the optical into the near-IR,
were detected and compared to synthetic spectra of model atmosphere
simulations (e.g. Stephens et al. 2009; Witte et al. 2011, Patience et
al. 2012).  Researchers are keen to reproduce both observed spectra and also
each others model results, leading to dedicated bench mark efforts for
example for dust cloud models (e.g. Helling et al. 2008). More often,
we learn something new only if model simulations do not fit
observations. Jones \& Tsuji (1997) compared their static model
atmosphere results to late M-dwarf spectra. The synthetic spectra only
started to be comparable to observations when the authors reduced
individual element abundances artificially, arguing these element
would be locked in dust grains and, hence, be not available to the
formation of molecules. Saumon et al. (2006) showed that the {\it
Spitzer} observation of ammonia (NH$_3$) indicates vertical mixing of
hotter material into detectable layers, hence, a local chemical
dis-equilibrium. Their chemical equilibrium model did not fit the
observations unless they artificially reduced the NH$_3$ abundances,
arguing that the reaction timescale of N$_2$ to NH$_3$ is much slower
than the convective mixing timescale.

Clearly, clouds play an important role in every atmosphere where they
are forming because they consume elements, and by this, change the local
gas-phase chemistry. Cloud particles have large radiation absorption cross
sections and they therefore increase the greenhouse effects, hence
affecting the local temperature. Furthermore, these cloud form at a
highly convective environment which drives a vivid turbulence field
that can initiate dust formation (Helling et al. 2001), and which increases
relative velocities between grains. Clouds have been observed to produce
discharge events like lightning and sprites in planet of our solar system
that carry clouds. Therefore, we have good reasons to expect
that cloud-forming extrasolar planets and brown dwarfs show
similar electrostatic activities.

We summarise our model of mineral cloud formation
(Sect.\,\ref{s:CloudModel}) and discuss in Sect.\,\ref{ss:thund} if
mineral clouds could produce lightning-like discharge events.
Section\,\ref{s:clio} describes collisional ionisation and ionisation by
cosmic rays as sources of charge separation in mineral clouds.

\section{Mineral cloud formation in extrasolar planets and brown dwarfs}\label{s:CloudModel}
Brown dwarfs and gas giant planets outside of our solar system are
likely not to form cloud made only of liquid droplets, but their
warmer atmospheres do allow solid dust particles to condense from the gas phase.

\begin{figure}[!h]
{\ }\\*[-0.1cm]
\centerline{\includegraphics[width=\columnwidth]{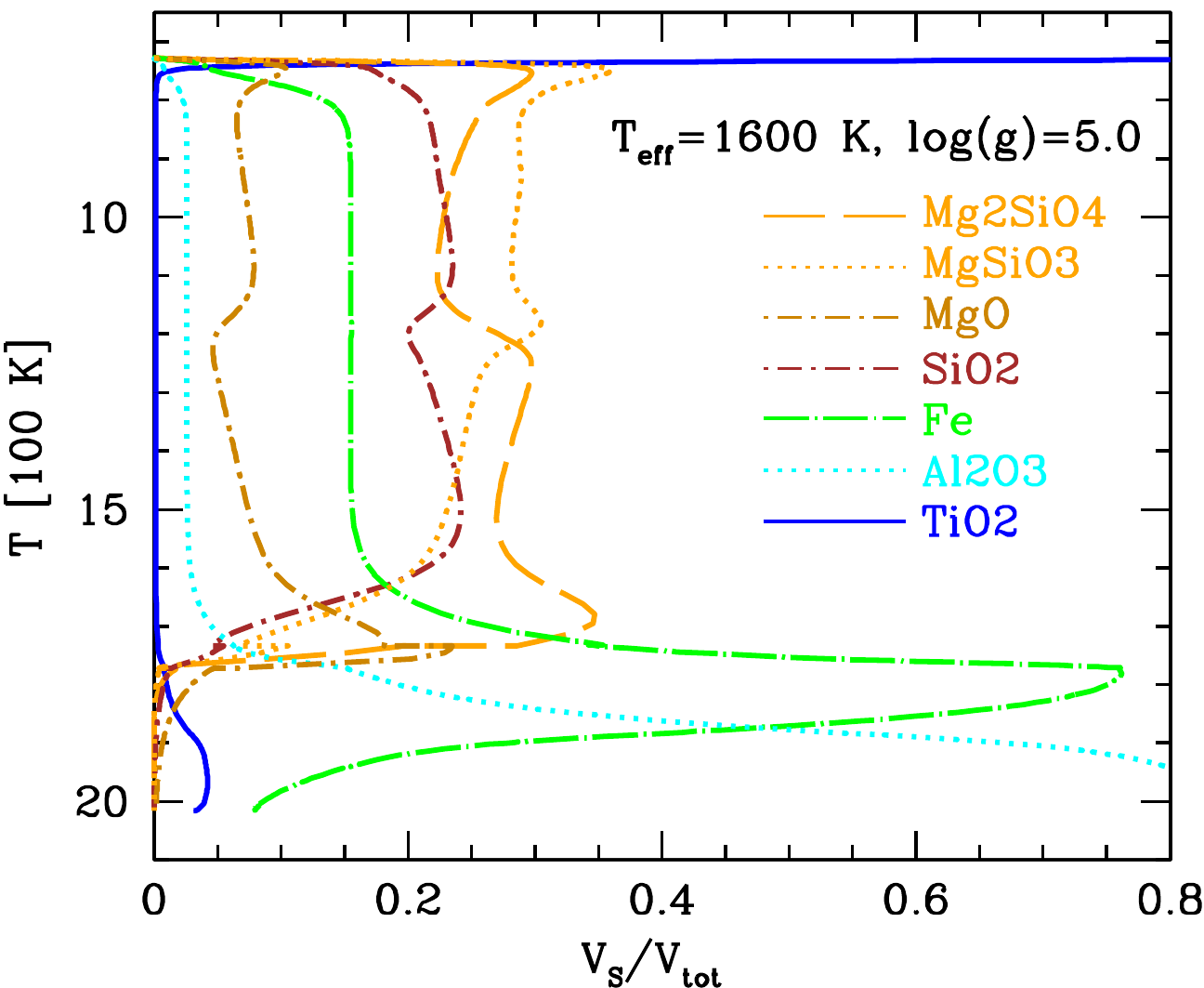}}
\caption{Dust cloud material composition in volume fractions $V_{\rm s}/V_{\rm tot}$ in a giant gas planet atmosphere (Helling et al. 2008) as result of {\sc Drift-Phoenix} model atmosphere simulations 
that include our kinetic dust formation model (Witte et al. 2009; T$_{\rm eff}$ - effective temperature of object, log(g) - surface gravity of object).  The composition changes with atmospheric height indicated by the local temperature. }
\label{VF}
\end{figure}

We have modelled the formation of such mineral clouds by describing
seed formation (by homogeneous nucleation) followed by the growth of 13
silicate and oxide solids by 60 chemical surface reactions,
evaporation, gravitational settling (rain out), convective element
replenishment, and element conservation (Woitke \& Helling 2003, 2004;
Helling \& Woitke 2006 Helling et al. 2008b). Our model
calculations start from solar element abundances which are
subsequently depleted by seed formation and the growth of the grain
mantle. If the grains become thermally unstable and evaporation sets
in, the element abundances will be enriched by those elements
previously locked in grains. Both processes, element depletion and
element enrichment, are non-uniform and individual for each involved element.
Processes between dust particles that lead to a further increase
in grain size, like for example coagulation, are not part of our
kinetic dust model because coagulation acts at a much longer time
scale. Coagulation is about 100$\times$ slower than the growth process
by surface reactions, hence, the formation processes (nucleation and
growth) will be much faster (Helling et al. 2008a).

The onset of dust formation is triggered by a nucleation process that
strongly depends on the local gas temperature and a high
supersaturation of the seed forming gas species which requires a
temperature well below thermal stability.  Typical supersaturation
ratios of our nucleation species TiO$_2$ are well above $10^4$ in the
nucleation region of the cloud (Fig. 1 in Helling et al. 2008b). The onset of growth requires the growing material to be
thermally stable only. The growth rate is determined by the inflow of
the growing gas phase constituents, hence, it is proportional to the
number density of the (grow-) contributing species and their velocity
distributions.  We refer to Helling \& Rietmeijer (2009) and above
mentioned papers for more details regarding the model equations. Our
solution of the kinetic dust formation predict a cloud structure as
function of the local gas temperature and gas density, $T$ and
$\rho_{\rm gas}$. Our model predicts   the mean grain size
$<\!\!a\!\!>(T,\rho_{\rm gas})$ [$\mu$m], the number density of dust
particles $n_{\rm d}(T,\rho_{\rm gas})$ [cm$^{-3}$], and the mean material
composition of the cloud particles $V_{\rm s}/V_{\rm tot}(T,\rho_{\rm
gas})$ [\%] (e.g. Fig.~\ref{VF}). We also calculate the chemical composition of the gas phase,
including the degree of ionisation (Sect.~\ref{ss:thund}). $V_{\rm s}$ is the dust volume occupied by the solid species s, $V_{\rm tot}$ is the total dust volume. To some extent,
the mean grain size, number of dust particles, the total dust volume, and higher dust moments
allows us to reproduce the grain size distribution, $f(V,
T,\rho_{\rm gas})$  which provides the number of dust grains for each grain volume $V$.

Clouds in brown dwarfs and extrasolar giant gas planets are composed
of a mixture of minerals due to the richness of the
atmospheric precursor gas in these objects (Fig.~\ref{VF}),
henceforth called \lq mineral clouds\rq.  The dust formation process (seed
formation, growth/evaporation) is influenced by gravitational
settling, hence, particle growth speeds up while the grains fall
inward along a positive density gradient. During this descent, the
crystal structure of the cloud particles is evolving (Helling \&
Rietmeijer 2009).  Figure~\ref{VF} indicates that such clouds are made
of small ($10^{-2}\mu$m) silicate particles at the top which develop
into large ($10\ldots 100\mu$m) iron/TiO$_2$ particles. For details on grain sizes see e.g.  Fig. 8 in Helling et al. (2008b).

Would dust-dust collisions change this picture?  The most interesting
changes in the grain size distribution by a dust-dust collision likely result from
the fragmentation of both collisional partners and the stick-and-hit events of projectile and target. 
Fragmentation would increase the number of grains and
therewith the number of seeds for further grow.  As surface growth is
rather efficient, the grain fragments can grow rather quickly to their
previous sizes until the gas-phase is undersaturated. Hence,
dust-dust collisions tend to increase the number of grains but the
grain size might not change as long as surface growth is efficient.
Stick-and-hit events would produce a higher number of large grains, but the collision energetic need to be just right.

\section{Sources of mineral dust cloud ionisation}\label{s:clio}
The reasons for ionisation in clouds are rather diverse, and more
complex processes than thermal ionisation need to be taken into
account because of the rapidly decreasing gas temperature with
height. They include energetic interstellar or interplanetary
radiation, radioactive decay, differences between surface potentials
of different materials (e.g. metals vs insulators), frictional
ionisation, collisional ionisation of accelerated charges, or
fragmentation of fractal particles (fracturing) . Triboelectric
charging is suggested to be of particular interest for dust cloud
charging on planetary surfaces (Sickafoose et al. 2001), a scenario
which is rather similar to mineral clouds in substellar atmospheres.
See also Saunders (2008) for an overview of the subject.  Also,
gas-phase ions can attach themselves to the grain surface, and by
this, contribute to the charging of cloud particles. Nicoll \&
Harrison (2010) have demonstrated this with observation of Earth
clouds.

\begin{figure}[!h]
{\ }\\*[-0.1cm]
\centerline{\includegraphics[width=10cm]{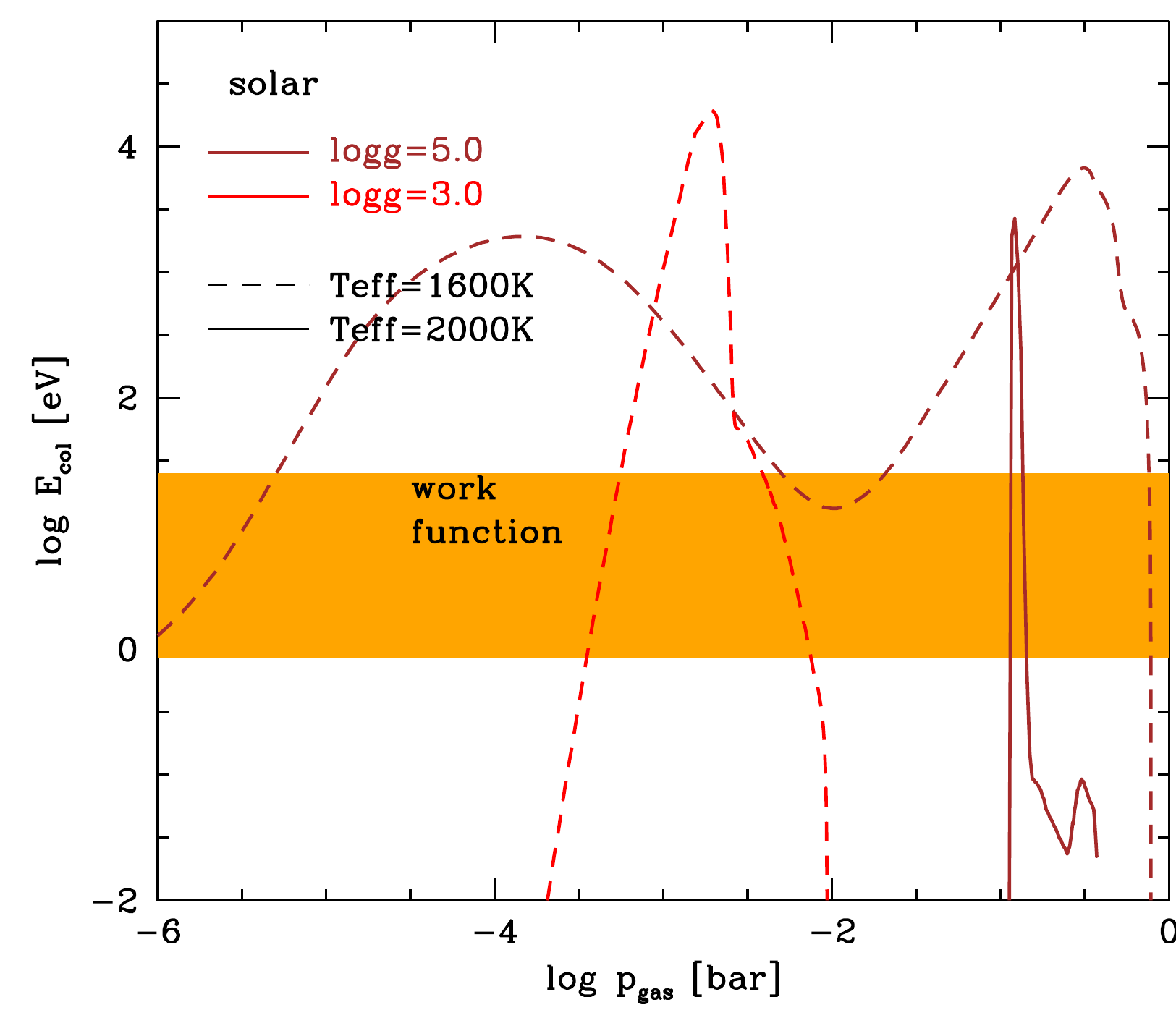}}
\caption{Turbulence enhanced dust-dust collision energies, $E_{\rm
col}$, for three different {\sc Drift-Phoenix} model atmosphere
simulations ($p_{\rm gas}$). The collision energy  is well above the work
function interval (orange bar) for the high-gravity, cool brown dwarf
(T$_{\rm eff}=1600$, K, log(g)=5.0) over a large pressure range in
contrast to its hotter counterpart (T$_{\rm eff}=2000$, K, log(g)=5.0).
}
\label{Ecol}
\end{figure}

 In this paper, we discuss two sources of dust ionisation that both result in the
ionisation of the cloud particles and eventually also in the ionisation
of the gas phase.

\begin{figure}
\begin{center}
\includegraphics[width=0.95\textwidth]{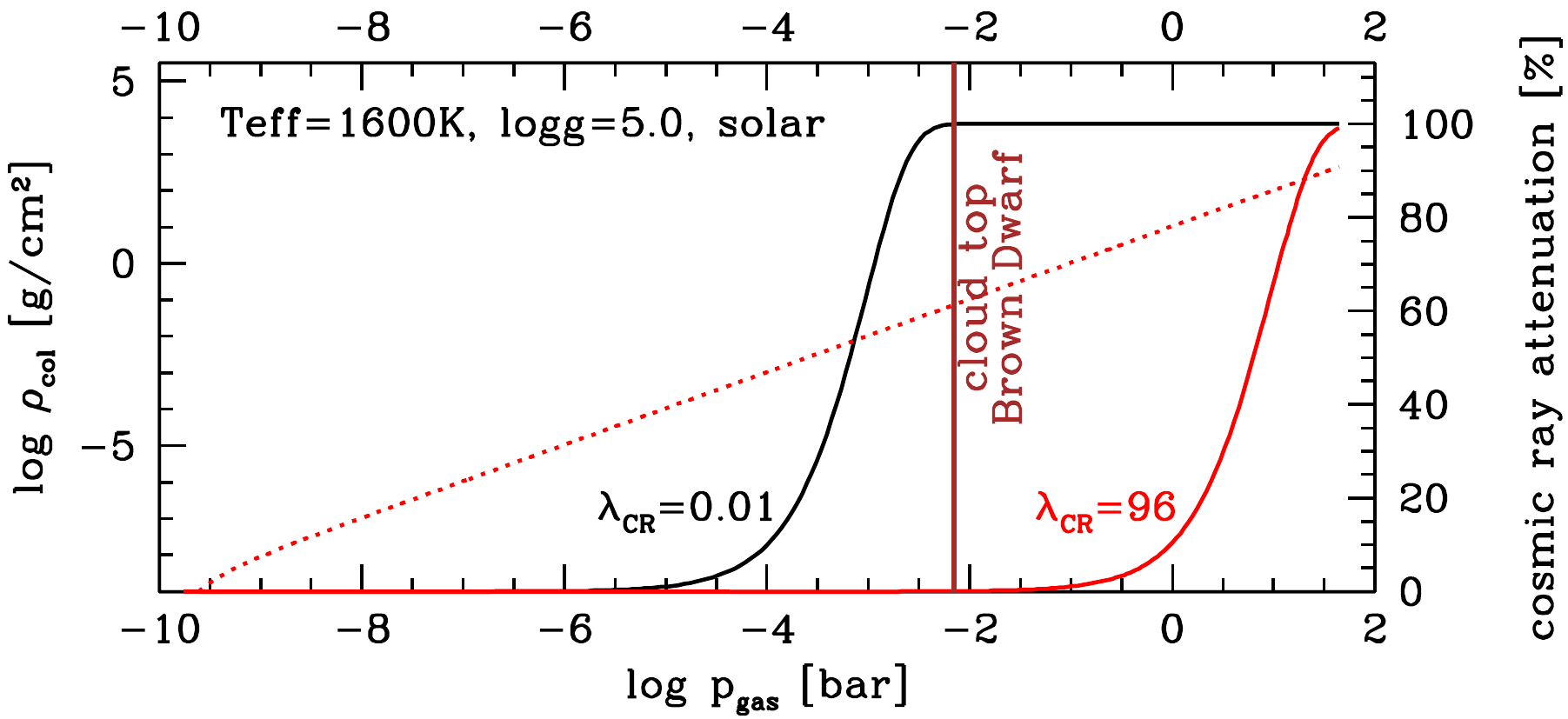}\\
\includegraphics[width=0.95\textwidth]{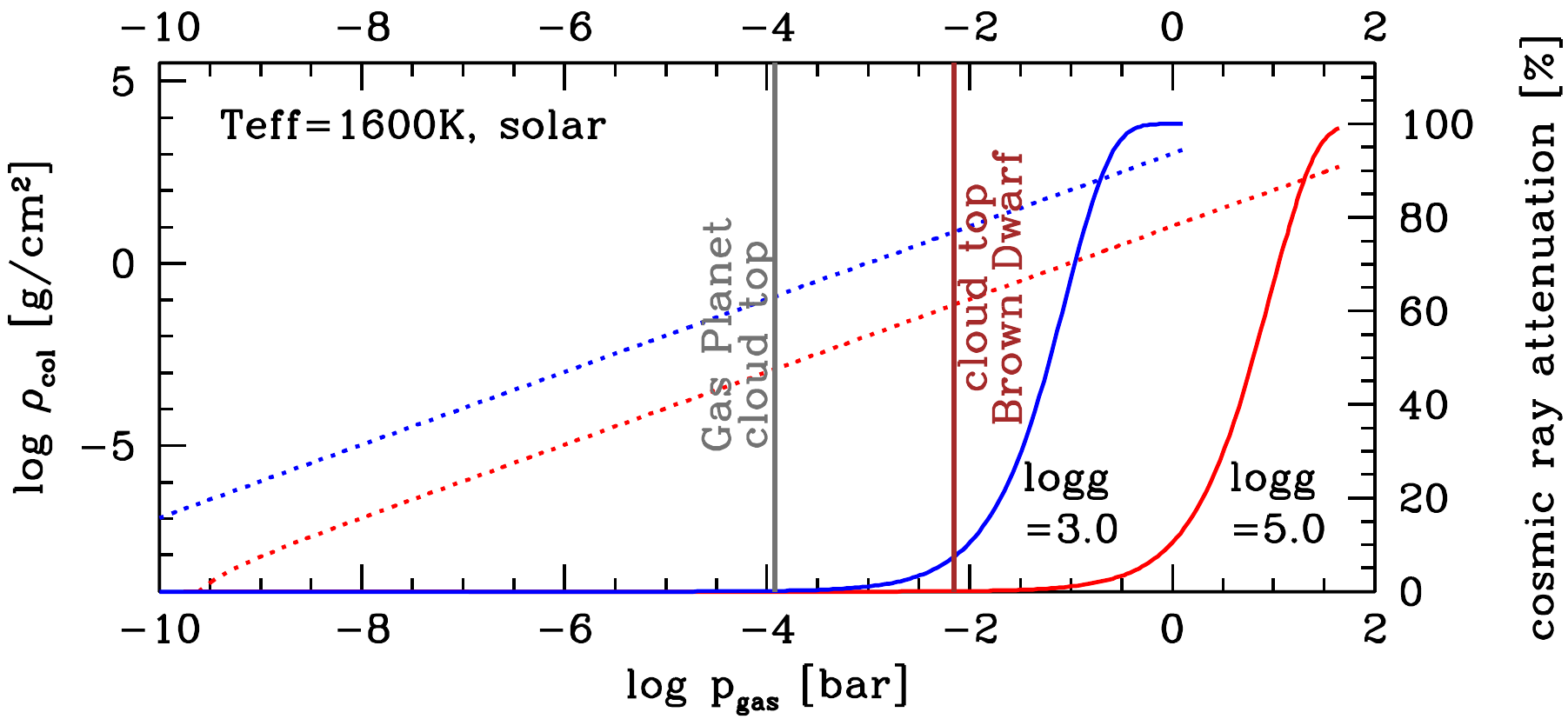}
\caption{\small Cosmic ray attenuation (in \%; solid lines, right
axis) and gas-phase column density  $\rho_{\rm col}$ (in g cm$^{-2}$; dotted lines,
left axis). {\bf Top:} The cosmic ray attenuation is shown for two
cases in a brown dwarf {\sc Drift-Phoenix} atmosphere (T$_{\rm
eff}=1600$K, log(g)=5.0, solar metallicity) with R$_*=10^9$cm =
0.14R$_{\rm Jupiter}$.: $\lambda_{\rm CR}=96$g cm$^{-2}$ (red solid line)
where H$_2$ provides the majority of the absorbing mass (Umebayashi \&
Nakano 2009) and $\lambda_{\rm CR}=0.01$g cm$^{-2}$ for which CR
attenuation reaches 100\% above the cloud top  above which grains
do form, too, but at a much lower rate.  $\lambda_{\rm CR}$ is the characteristic CR penetration length. {\bf Bottom:} Same as above but comparing a brown dwarf and a gas planet atmosphere. }
\label{fig:cra}
\end{center}
\end{figure}

\subsection[]{Dust-dust collisions}

Helling et al. (2011b) studied dust-dust collision kinematics
using results of our kinetic dust formation model from the model
atmospheres code {\sc Drift-Phoenix} (Dehn 2007, Helling et al. 2008b,
Witte et al. 2009). This collisional charge process is know as tribo-electric
charging and is suggested to work well for dust charging of Martian dust (Sickafoose et al. 2001).  Our results have shown that dust-gas and dust-dust collisions due
to gravitational settling (drift, rain out) are not energetic enough
to overcome the energy needed to release an electron from the crystal
structure (work function). Only turbulence-enhances dust-dust
collisions such that tribo-electric charging is possible.

Figure~\ref{Ecol} shows the turbulence-enhanced energy by dust-dust
collisions for three different atmosphere simulations.  This rather limited number of
{\sc Drift-Phoenix} investigated atmosphere models suggests that the
pressure range affected by tribo-electric charges is larger in a
compact, low temperature brown dwarf (T$_{\rm eff}$=1600K, log(g)=5.0)
compared to a hot brown dwarf (T$_{\rm eff}$=2000K, log(g)=5.0) or a
giant gas planet (T$_{\rm eff}$=1600K, log(g)=3.0). The geometrical
extension of the clouds, however, differ widely between brown dwarfs
and giant gas planets. Brown dwarfs have much less extended clouds
then giant gas planets due to their much higher surface gravity, and
hence smaller pressure scale height. This suggests a smaller
atmospheric volume being affected by charged mineral clouds in brown
dwarfs compared to Jovian plants.  However, our comparison of
streamer timescale and Coulomb recombination time scale with the time
scale on which dust particles pass through a previously formed
electron cloud (streamer), and hence, potentially initiate another
electron avalanche, suggest a stronger  
lightning activity in higher density environments such as brown dwarf
atmospheres.

\subsection[]{Cosmic ray attenuation}\label{ss:cra}

Cosmic rays (CR) appear to be an important source of atmospheric
ionisation in the solar system planets.  Observations of Earth clouds,
however, suggest that the actual charge production is not overly
efficient but potentially important for coagulation processes. Nicoll \&
Harrison (2010) determine a maximum mean droplet charge of 17e at the
cloud edges on Earth which can be directly related to ionisation by
cosmic rays. The charging of these water cloud particles is, however,
not a direct result of the impact of the high energy CRs on the cloud
but rather of the ion and electron currents that develop from the CR ionisation of
the gas above the cloud. These charges attach to the cloud particles.  A
similar scenario can be envisioned for close-in exoplanets that form
an ionosphere due to the X-ray and the extreme UV radiation of the
host star as demonstrated by Koskinen et al. (2010).

The question is whether galactic cosmic rays could be a global source
of ionisation for the whole atmosphere of extrasolar low-mass objects
which are not exposed to the high-energy radiation of a nearby host
star.  Ionisation by CRs includes the effect of
secondary particles like electrons but also $\gamma$-rays.  Umebayashi
\& Nakano (2009) show for protoplanetary disks that the CR attenuation increases
exponentially with a characteristic length or column density
$\lambda_{\rm CR}\sim 96$g cm$^{-2}$ of a gas where H$_2$ provides the
majority of the mass. This translates into a simple ansatz for the
attenuation,
\begin{equation}
\label{eq:attenu}
\frac{F(z)}{F_0}=\exp\big(-\frac{\rho_{\rm col}}{\lambda_{\rm CR}}\big),
\end{equation}
 with $F(z)$ the local CR flux at height $z$, $F_0$ the incident CR
flux, and $\rho_{\rm col}$ the gas column
density. Figure~\ref{fig:cra} shows $F(z)/F_0$, the attenuation of
cosmic rays (solid line, rhs. axis of figure), which is determined by
the local atmospheric gas-phase column density $\rho_{\rm
col}=\int_{z_0}^{z_{\rm i}} \rho_{\rm gas}(z)\,dz$ [g/cm$^2$] (dotted
line, lhs. axis of figure) integrated over the atmosphere extension
$z_0\,\ldots\,z_{\rm i}$. The attenuation increases inward (downward)
with increasing gas-phase column density. 

Cosmic rays penetrate the cloud almost unhindered if H$_2$ is the only
absorber ($\lambda_{\rm CR}= 96$g cm$^{-2}$ in
Fig.~\ref{fig:cra}). The gas column density is high enough for
complete attenuation only at very high gas pressures in both the brown
dwarf and the gas giant.  Brown dwarfs, on the other hand, have been
shown to maintain magnetic fields (e.g. Reiners \& Basri 2008;
Hallinan et al. 2008; Christensen, Holzwarth \& Reiners 2009; Berger
et al. 2010) and also planets have a magnetosphere (e.g. Zarka et
al. 2001, Jardine \& Cameron 2008). Dolginov \& Stepinski (1993) argue
that the magnetic field will diffuse the cosmic rays paths through the
atmosphere and, hence, decrease the critical column density (or scale
height) $\lambda_{\rm CR}$ for cosmic ray attenuation. Glauser et
al. (2009) demonstrate that fast protons and He ions interact with
dust grains in disks, hence, the dust will decrease the effective
attenuation length in the atmosphere further.  This suggests that a
smaller fraction of the atmosphere and of the cloud is influenced by
cosmic rays. Nontheless, the critical column density needs to be as low as
$\lambda_{\rm CR}=0.01$g cm$^{-2}$ (Fig.~\ref{fig:cra}, black solid
line) to achieve 100\% attenuation above the cloud in the brown dwarf
model atmosphere studied here. It is, therefore, not clear if cosmic
rays can be a global source for cloud ionisation as brown dwarfs have
magnetic fields which are particularly strong on small scales. The
B-field shielding would, however, be less efficient in the polar
regions if the magnetic field is predominantly dipolar, an assumption
which is justified for low-mass objects (see Morin et al. 2010).  The consequence would be an increased polar discharge activity which is triggered by an increased degree of cloud ionisation due to CRs.

\section{Should we expect thunderstorms in mineral dust clouds?}\label{ss:thund}

\begin{figure}[!h]
{\ }\\*[-0.1cm]
\centerline{\includegraphics[width=10cm]{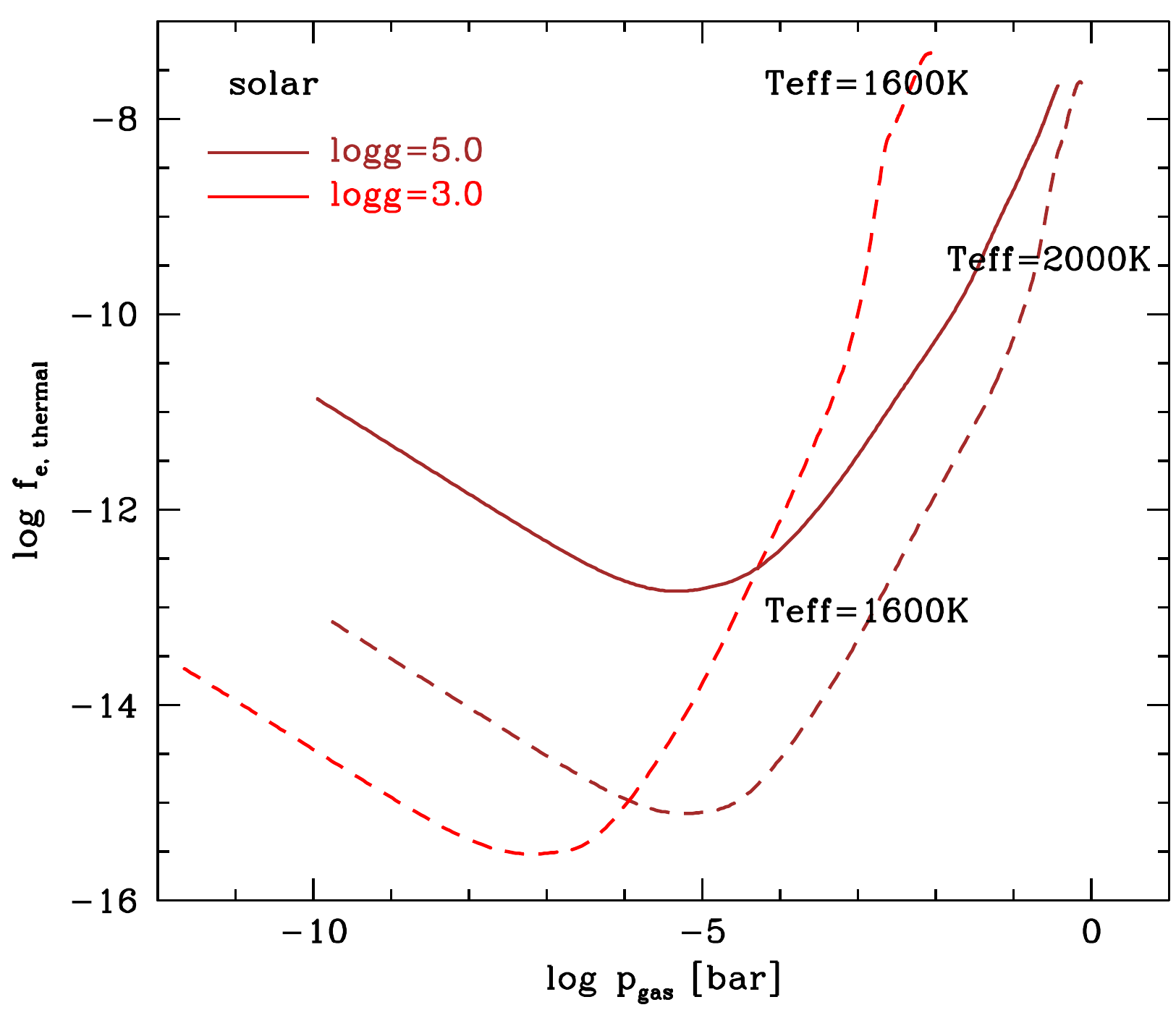}}
\caption{Degree of thermal gas-ionisation, $f_{\rm e, thermal}=p_{\rm e,
thermal}/p_{\rm gas}$,  in the cloud for the same three example model atmospheres like in Fig.~\ref{Ecol}. (T$_{\rm eff}$ - effective temperature of object, log(g) - surface gravity of object).}
\label{fe}
\end{figure}

Brown dwarf atmospheres are the perfect example for a neutral
atmosphere if thermal ionisation is considered only (Mohanty et
al. 2002). Figure~\ref{fe} shows the degree of thermal ionisation of
the gas phase in the pressure interval where the cloud forms for
three different atmosphere models. The degree of ionisation increases
inward towards higher temperature due to thermal ionisation but drops
rather quickly in the low pressure regime. This figure demonstrates
that the degree of thermal ionisation, $f_{\rm e, thermal}=p_{\rm e,
thermal}/p_{\rm gas}$, is less than $10^{-8}$.  It is important to
understand that although thermal ionisation does not provide enough free
charges to produce, for instance, a magnetic Reynolds number $>$ 1,
which would indicate the atmospheres potential for coupling to the
strong magnetic field of brown dwarfs, the number of free electron
may be large enough to produce a streamer between two charged dust
grains or a whole front of charged dust grains (Dowds et al. 2003, Helling et al. 2011a,b)

\begin{figure}[!h]
{\ }\\*[-0.1cm]
\centerline{\includegraphics[width=10cm]{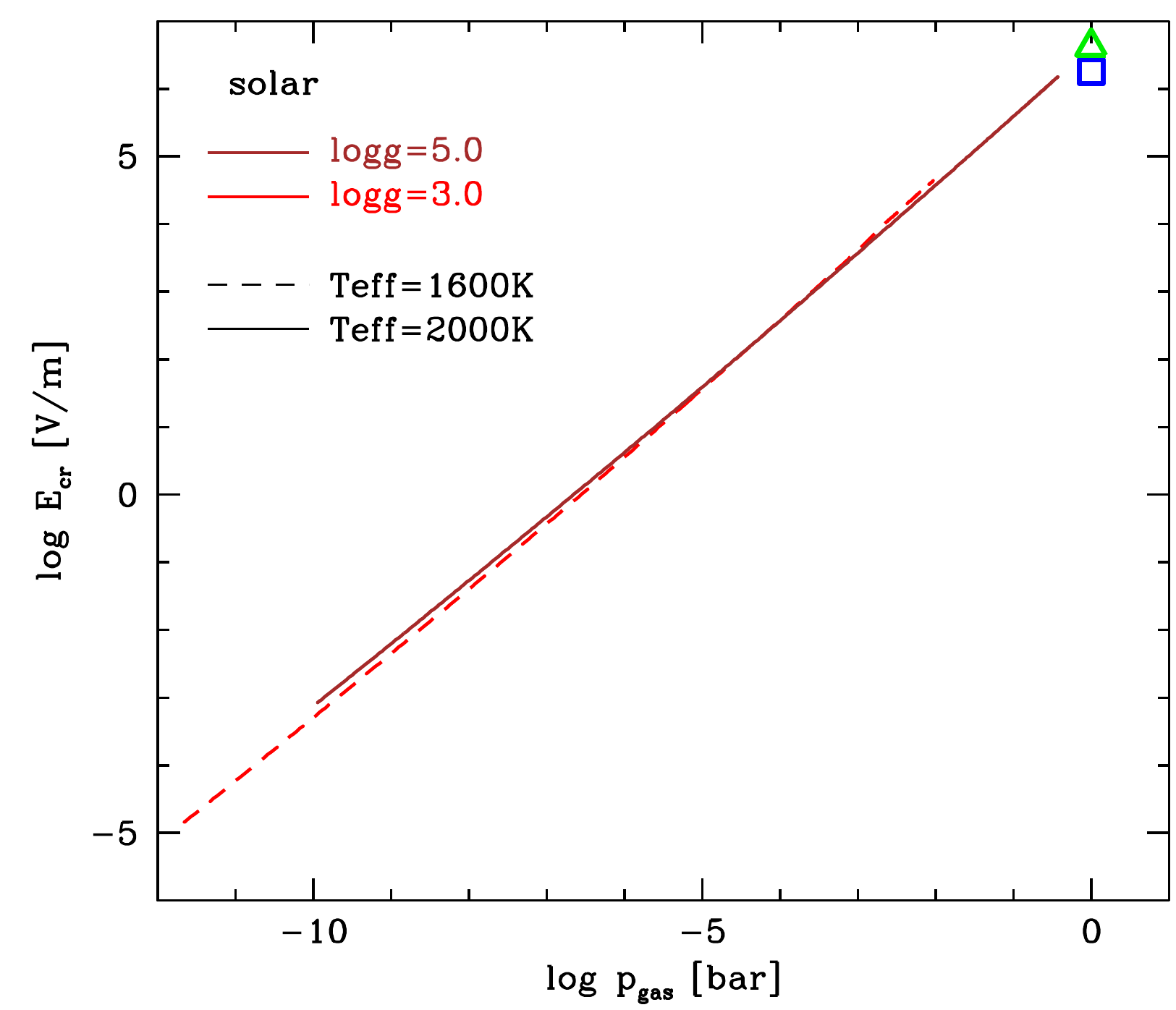}}
\caption{The break-down
field strengths, $E_{\rm crit}$, for three different brown dwarf and
giant gas planet {\sc Drift-Phoenix} model atmosphere of an initial
solar element composition: $T_{\rm eff}=1600$K, log(g)=5.0 (brown
dashed line; brown dwarf), $T_{\rm eff}=1600$K, log(g)=3.0 (red dashed
line; giant gas planet), $T_{\rm eff}=2000$K, log(g)=5.0 (brown solid
line; brown dwarf). The break-down field is calculated for a
H$_2$-rich gas in the model atmosphere simulations of initial solar element abundances. Values for Earth
(green triangle) and Jupiter (blue square) at 1 bar are shown for
comparison.}
\label{Ecr}
\end{figure}

Figure~\ref{Ecr} shows the electric field that needs to be overcome by
a mineral cloud to initiate a lightning-discharge. The values for
Earth (green triangle) and for Jupiter (blue square) are shown for
comparison and pressure unit check. Note that these values follow the
classical break-down field parameterisation as given in Yair et
al. (1995) but that field measurements above thunderclouds suggest an
electric break-down field of 1-2 orders of magnitude less. Clearly,
the break-down is much easier in low-pressure regimes which is
apparent from Fig.~\ref{Ecr} for the studied model atmospheres. Note,
however, that the Paschen curves shown in Fig.~\ref{Ecr} would turn
into the Paschen minimum if continued to lower pressures before they
become unphysically asymptotic at very low pressures.

For a given chemical composition, the break-down field is a strong
function of gas pressure, therefore it is of similar orders of
magnitude for all models which differ in gravity and effective
temperature. All model simulations are performed for a
hydrogen-dominated gas, with H$_2$ being the main ionised
species. Hydrogen remains the most abundant molecule also in the cloud
forming part of the atmosphere where far less abundant elements
(e.g. Si, Fe, Mg, O) are depleted by dust formation (Fig. 4 in Helling
et al. 2008a).
 
\begin{figure}[!t]
{\ }\\*[-0.1cm]
\centerline{\includegraphics[width=10cm]{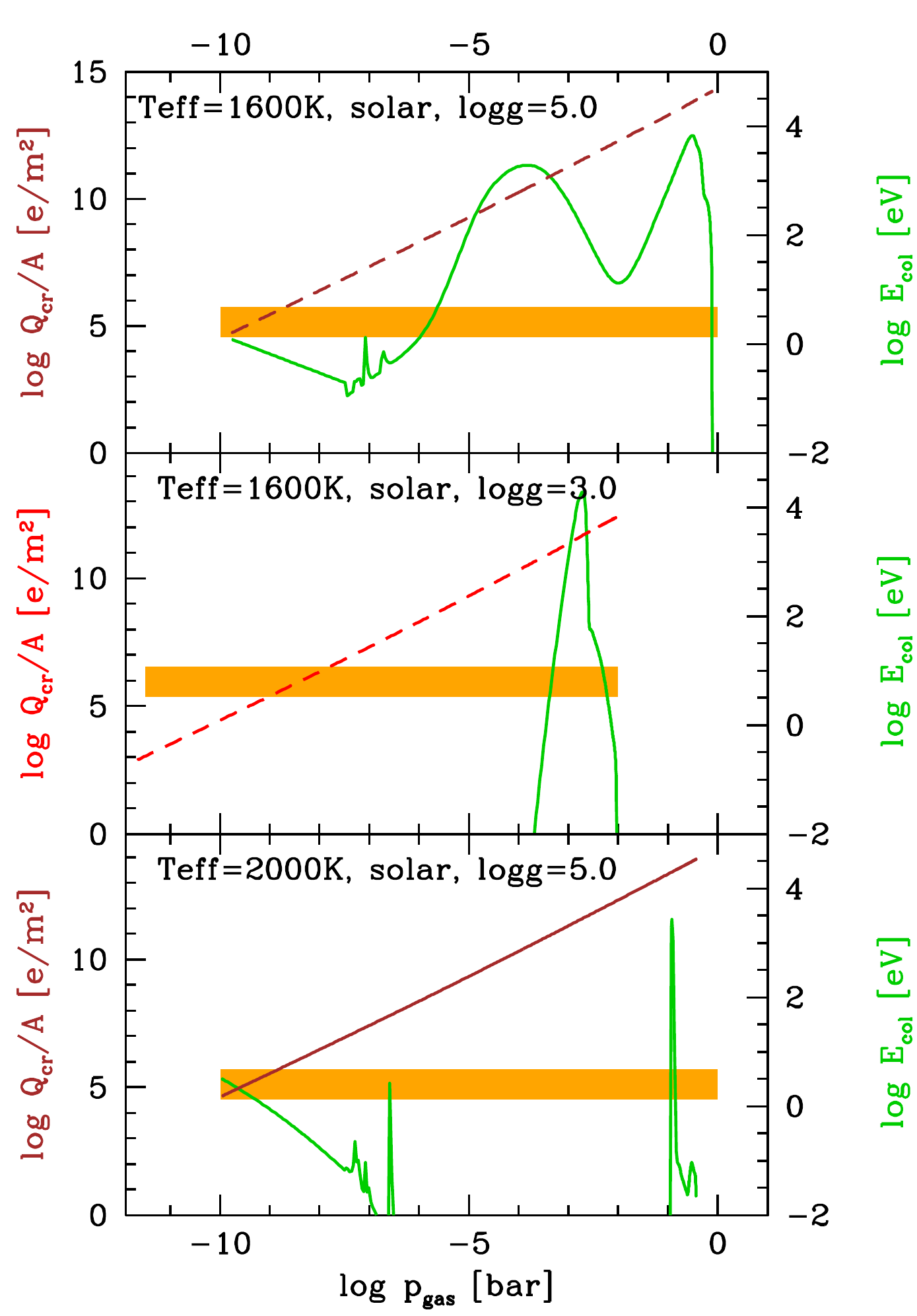}}
\caption{Number of charges per m$^2$ (red and brown lines; left axis)
needed to produce the break-down field shown in Fig.~\ref{Ecr}. The
overplotted (green; right axis) lines is the energy produced during
dust-dust collisions as in Fig.~\ref{Ecol}. The orange bar indicates the interval of work functions of different materials. This figure demonstrate where dust-grains can be charged by intra-cloud collisional processes
compared to the distribution of needed break-down charges inside the
cloud.  Note that the geometrical extension of the cloud differs between brown dwarfs and gas giants which is not reflected by the pressure scale shown in this plot. The comparison is shown for the same three atmosphere models like in Fig.~\ref{Ecr}.}
\label{Qperm2}
\end{figure}

We conclude that mineral clouds should also be able to produce
 discharge events as the break-down field are considerably lower than
 on Earth and Jupiter ($\sim 10^6$ V/m) in the cloud forming
 regions. We suggest that lightning discharges should be
 expected in the upper part of the atmosphere and the mineral cloud,
  as gravitational settling provides a mechanisms for large-scale
 charge separation. Particles of different sizes fall with different
 velocities which leaves the smaller, less negatively charged
 (e.g. Merrison et al. 2012) cloud particles suspended for longer in
 the upper cloud layers.  We reached a similar conclusion in a
 previous paper (Helling, Jardine \& Mokler 2011) where we compared
 streamer timescales and Coulomb recombination time scales with the
 time scale on which dust particles pass through a previously formed
 electron cloud (streamer) and, hence, potentially initiate further
 electron avalanches. Our understanding is that a superposition of
 avalanche-streamer processes will lead to more and more free
 electrons for a short time period which then may be defined as
 lightning. Such a superposition is more likely in the upper,
 low-pressure part of the cloud, too.
 
\medskip  
How much charges do we need to achieve an electric field break down?
Treating a cloud as a capacitor allows to approximate the number of
charges per m$^2$, $Q_{\rm cr}/A = E_{\rm cr}\cdot\epsilon_0$
($\epsilon_0$ - electric constant), needed to overcome the break-down
field. Figure~\ref{Qperm2} shows the number of charges per surface
area needed to overcome the break-down field is strongly
height-dependent in the atmosphere.  The atmospher's height is
represented by the pressure scale in Fig.~\ref{Qperm2}: high pressures
refer to lower atmospheric heights with high densities, and low
pressures refer to the higher atmospheric layers with low
densities. The number of charges per surface area in Fig.~\ref{Qperm2}
is compared with the pressure regime in which dust-dust collisions are
most likely producing charged dust particles. Our results suggest that
dust plays an important role in producing free charges in most of the
cloud of cool brown dwarfs, for example by triggering electron
avalanches. Only a fraction of the cloud can be charged by dust-dust
collisions in hotter brown dwarfs and in the less-dense giant gas
planet atmospheres. Note, however, that the cloud in a giant gas
planet is $100\times$ more extended than the clouds forming in brown
dwarf atmospheres with high surface gravity.

\section{Conclusions}
Assuming that cloud particles are charged in brown dwarf and
exoplanetary atmospheres, then electron avalanche processes are
initiated between two charged grains and develop to a streamer's
ionisation front (Helling et al. 2011a).  We have argued that a large
part of the clouds in brown dwarfs and extrasolar planets is
susceptible to local discharge events which are triggered by charged
dust grains.  Such discharges occur on time scales shorter than the
time required to neutralise the dust grains, and their superposition
might produce enough free charges to suggest a partial and stochastic
coupling of the atmosphere to a large-scale magnetic field. Discharge
processes in brown dwarf and exoplanetary atmospheres should not connect
to a crust as on terrestrial planets, hence, they will experience
intra-cloud discharges comparable to volcano plumes and dust devils.

\medskip
{\bf Acknowledgement:} ChH acknowledges an ERC starting grant for the
LEAP project of the EU program FP7 {\it Ideas}. Most of the literature
search was done with the ADS. The computer support at our institutes
is highly acknowledged.

\bibliographystyle{elsarticle-harv}
\bibliography{<your-bib-database>}



\end{document}